\documentclass[final,5p,times,twocolumn]{elsarticle}

\usepackage{graphicx}

\usepackage[english]{babel}
\usepackage[utf8]{inputenc}
\usepackage{epstopdf}
\usepackage{graphicx}
\usepackage{caption}
\usepackage{subcaption}
\graphicspath{{figures/}}
\usepackage{booktabs}
\usepackage[exponent-product = \cdot]{siunitx}
\usepackage{multirow}
\usepackage{lmodern}
\usepackage{caption}
\usepackage{siunitx}
\usepackage[version=4]{mhchem}
\usepackage{physics}
\usepackage{bm}
\usepackage{tikz-dimline}
\usepackage{adjustbox}
\usepackage{overpic}
\usepackage{url}
\usepackage{tikz}
\usetikzlibrary{decorations.pathreplacing,calc}
\usetikzlibrary{positioning}
\usetikzlibrary{arrows.meta}
\usetikzlibrary{decorations.markings}
\usetikzlibrary{patterns}
\usetikzlibrary{shapes}
\usetikzlibrary{3d}
\usetikzlibrary{backgrounds}
\usetikzlibrary{fit}
\usetikzlibrary{matrix}
\usepackage{tikz}
\usetikzlibrary{arrows.meta}
\usetikzlibrary{positioning}
\usetikzlibrary{shapes.geometric}
\usetikzlibrary{calc}

\usepackage{pgfplots}
\usepgfplotslibrary{groupplots}
\usepgfplotslibrary{colormaps}
\usepgfplotslibrary{colorbrewer}
\pgfplotsset{cycle list/Dark2}
\pgfplotsset{compat=newest,/pgf/number format/.cd,1000 sep={}}
\pgfplotsset{colormap={modrainbow}{rgb255(0cm)=(0,0,255); rgb255(1cm)=(0,255,255); rgb255(2cm)=(0,255,0);
rgb255(3cm)=(255,255,0); rgb255(4cm)=(255,0,0)}}
\usepackage{wasysym,amssymb}

\journal{Journal of Computational Physics}

\begin{document}

\begin{frontmatter}



\title{Reactive Trace Species Modeling with the Simulation Tool PICLas}
	\author[a]{S. Lauterbach}
    \author[a]{K.-S. Ellenberger\corref{cor1}}
	\ead{ellenbergerk@irs.uni-stuttgart.de}
	\author[a]{S. Fasoulas}
	\author[a]{M. Pfeiffer}

    \cortext[cor1]{Corresponding author}

	\affiliation[a]{organization={Institute of Space Systems, University of Stuttgart},
	             addressline={Pfaffenwaldring 29},
	             city={Stuttgart},
	             postcode={70569},
	             country={Germany}}


%\tnotetext[tn1]{Preprint submitted to Journal of Computational Physics.}


% ----------------------------------------------------------------------------------------------- %
% Abstract
% ----------------------------------------------------------------------------------------------- %
\begin{abstract}
A species-specific weighting scheme for the Natural-Sample-Size collision model within the Direct Simulation Monte Carlo method is developed and implemented in the open-source gas and plasma solver PICLas. The implemented weighting scheme places particular emphasis on chemical reactions and ionization processes of trace species in order to guarantee physically accurate and low-noise simulations of atmospheric entry processes. The successful implementation and the conservation of momentum and energy during relaxation processes and chemical reactions are verified through a reservoir simulation of a reacting air mixture. As a more sophisticated validation case, atmospheric entry simulations of a cylinder at an altitude of 108 km are investigated. In a final step, the developed weighting scheme is employed to investigate ionization processes in the shock layer surrounding the RAM-C II reentry probe during atmospheric entry. The simulations showed very good agreement with experimental flight data and reduced noise in the rarefied regions for the charged species.
\end{abstract}

% ----------------------------------------------------------------------------------------------- %
% Graphical abstract, research highlights, and keywords
% ----------------------------------------------------------------------------------------------- %

% \begin{highlights}
%     \item Development of advanced particle weighting strategies for trace species
%     \item Special consideration of reactive and ionization processes
%     \item Validation for atmospheric entry simulations
%     \item Accurate reproduction of flight-measurement data for the RAM-C reentry probe
% \end{highlights}

%% Keywords

\end{frontmatter}

\section{Introduction}

During hypersonic entry into the rarefied upper atmosphere, the space vehicle experiences extreme conditions, such as high temperatures and pressures in the shock layer, which can lead to the occurrence of gas-phase chemical reactions and favor ionization of atmospheric species. Under these conditions, even processes involving components in relatively low number densities can govern the overall flow physics by the cumulative effects of the trace species. Trace chemical reactions can result in the formation of highly energetic and reactive species that affect the shock-layer structure and the surface properties of the entry vehicle. Furthermore, even small degrees of ionization may lead to the formation of a plasma and thus interfere with radio-frequency signal transmission, leading to communication blackout phenomena \cite{takahashi2016analysis}. Critical properties for the prediction of communication blackout, such as the electron temperature, are not directly measurable by in-flight experiments. Therefore, accurate numerical predictions of the underlying trace effects are required for consideration in design processes of entry vehicles.\\
The Direct Simulation Monte Carlo (DSMC) method \cite{bird1994molecular} has been widely established for the simulation of rarefied, non-equilibrium flows. However, the particle-based nature of the method poses a challenge for the efficient handling of large variations between species densities, as commonly encountered in ionizing hypersonic flows. In standard DSMC implementations, the use of a single, uniform particle weight either results in prohibitively large particle numbers for the denser species or in high statistical noise and insufficient sampling accuracy for the trace components. With the use of species-specific weighting factors, trace species can be represented with significantly higher particle numbers without a substantial increase in the overall computational demand.
However, the treatment of reactive processes under conservation of mass, momentum and energy remains challenging.
While previous developments of such weighting schemes have focused on the Constrained Probability (CP) collision model within DSMC \cite{boyd1996conservative,galitzine2015adaptive,fang2020dsmc,boyd2007modeling}, no such scheme has been considered for the Natural-Sample-Size (NSS) collision model yet. \\

In the present work, a species-specific weighting scheme is developed for the NSS collision model in the gas and plasma simulation tool PICLas \cite{fasoulas2019combining}. PICLas\footnote{github.com/piclas-framework/piclas} is an open-source code which is cooperatively developed by the Institute of Space Systems (IRS) at the University of Stuttgart and the spin-off company boltzplatz. The simulation tool combines different particle-based methods such as DSMC, with the objective of solving the Boltzmann equation.\\
In the following, the differences between the NSS and CP collision models within DSMC method are discussed first. Subsequently, the development and implementation of the novel species-specific weighting scheme are presented in detail, with a particular focus on the necessary modifications for chemical reactions. The functionality of the novel method, especially with regard to conservation properties in relaxation processes and chemical reactions, is verified by a reservoir simulation of a reactive air mixture. Furthermore, hypersonic entry simulations of a 2D cylinder at an altitude of 108 km are analyzed in detail. Here, a special consideration is given to the statistical behavior of the novel weighting scheme and the preservation of higher-order moments of the distribution function, as reflected by the evaluation of the temperature flow field and the surface heat flux.
Finally, the species-specific weighting scheme is applied to the simulation of the RAM-C II atmospheric entry experiments in order to investigate ionization processes within the shock layer and the simulated electron density along the probe surface is compared to available experimental flight data.
% at an altitude of \SI{81}{\kilo\meter}. The RAM-C II experiment has been previously used to verify several new weighting schemes, but with the CP and not the NSS collision model. Here, electrons in the flow are treated with a weighting factor scheme for trace species, a higher weighting factor overall, or a continuum approach at altitudes of \SI{73}{\kilo\meter} to \SI{81}{\kilo\meter} in 2D axis-symmetric or full 3D simulations\cite{fang2020dsmc,boyd2007modeling,shevyrin2020calculation}. \textcolor{red}{Oder hier ausführlicher?}

\section{Methods}

\subsection{Direct Simulation Monte Carlo Method}
 The physical description of the gas flow in PICLas is based on the Boltzmann equation, which describes the microscopic gas state by the change in the particle distribution function $\pdv{f}{t}$. This function in turn depends on the particle positions $\vec{x}$, the particle velocities $\vec{v}$, and the time $t$
     \begin{equation}
     	\pdv{f}{t} + \vec{v}\cdot\pdv{f}{\vec{x}} = \left. \pdv{f}{t}\right|_{\text{coll}}.\label{eq:boltzmann}
     \end{equation}
The collision integral on the right-hand side of the equation describes particle interactions in binary particle collisions. In this equation, external forces, such as electro magnetic fields, are neglected. As the integro-differential equation is difficult to evaluate analytically, the probabilistic DSMC method is employed to obtain an approximate solution. In the method, as first introduced by Bird \cite{bird1994molecular}, the particle distribution function is approximated by a finite number of simulation particles. In order to minimize the computational demand, each simulation particle typically represents a large number of real particles, as defined by the weighting factor $w$. In conventional DSMC implementations, a uniform weighting factor is employed for all species. \\ Particle motion and interaction steps are handled consecutively. During the movement step, the particle positions are updated based on Newton's equations of motion, while in the interaction step, particles are paired according to a nearest-neighbor scheme \cite{pfeiffer2013grid}, in combination with a recursive Octree algorithm. The different schemes for the statistical evaluation of particle collisions are detailed in the following subsection \ref{ssec:coll}. Depending on the collision energy and the species involved, collisions may result in the relaxation of internal energy modes for molecules or in chemical reactions between the collision partners. In the final step of the DSMC algorithm, macroscopic flow properties are sampled from the moments of the distribution function and averaged over all particles in a simulation cell.

\subsection{Collision Schemes}\label{ssec:coll}
As DSMC aims to reproduce the Boltzmann equation through the stochastic evaluation of binary particle collisions, the number of potential collision pairs $Q_{\text{AB}}$ and collision probabilities $P_c$ need to be chosen such, that the correct physical collision rate $Z_{AB}$ between the species $A$ and $B$ within a given volume and time frame, as predicted by the principles of gas kinetic theory, is reproduced:
\begin{equation}
    Z_{\text{AB}}=\langle P_c \rangle Q_{\text{AB}}.
\end{equation}
In DSMC simulations, the instantaneous collision probabilities $P_c$ are calculated for each collision pair and compared with a random number $R\in [0;1]$, which reproduces the correct average collision probability $\langle P_c \rangle$. Using the particle weighting factor $w$, the number of physical collision pairs $Q_{\text{AB}}$ is replaced by an interaction number in terms of simulation particle pairs
\begin{equation}
    S_{\text{AB}}=w Q_{\text{AB}}.
\end{equation}
This expression is only directly applicable when equal weights are employed for both species. Multiple approaches exist for the determination of $S_{AB}$ and $P_c$ in order to reproduce the real collision rate, as proposed by Baganoff and McDonald \cite{baganoff1990collision, mcdonald1990computationally}. While the CP approach is predominantly employed, the DSMC method in PICLas utilizes the NSS method.

\subsubsection{Constrained Probability}
In the CP method, the number of potential collision pairs is first calculated in each time step according to
\begin{equation}
    S_{AB} = \frac{N_A N_B}{1+\delta_{AB}}\frac{w\Delta t}{V}(\sigma_{AB} g_{AB})_{\text{max}}
\end{equation}
with the number of simulation particles $N$ for species $A$ and $B$ in the cell volume $V$. The Dirac delta function $\delta_{AB}$ is zero for collision pairs of unequal species and one for collisions between particles of the same species. Afterwards, the collision probability for each selected pair is determined by the collisional cross-section $\sigma_{\text{AB}}$ according to the Variable Hard Sphere (VHS) model, and the relative velocity $g_{\text{AB}}$ of the interacting particles
\begin{equation}
		P_{\text{c}}=\frac{\sigma_{AB} g_{AB}}{(\sigma_{AB} g_{AB})_{\text{max}}}.
\end{equation}
This way, the collision probability is always defined in a meaningful manner, with values in the range $[0;1]$. However, the maximum value of the collisional cross section $(\sigma_{AB} g_{AB})_{\text{max}}$ needs to be stored and updated dynamically for each simulation cell. Furthermore, a particle may undergo multiple collisions in a single time step. \\
Extensions of the CP method to handle collisions between differently weighted particles have been first proposed by Bird \cite{bird1994molecular} and further developed by Boyd and Galitzine \cite{boyd1996conservative,galitzine2015adaptive}. In these modified collision schemes, both the number of collision pairs and the velocity update scheme are adapted to reproduce the number of physical collision pairs and ensure momentum and energy conservation on average.

\subsubsection{Natural-Sample-Size}
In the NSS approach, all particles in a simulation cell are paired directly, such that $S_{\text{AB}} = N_{\text{cell}}/2$ potential collision pairs are considered in every time step. The collision probability for a given pair is then determined according to the following equation:
    \begin{equation}
        P_c = w_{\text{AB}}\frac{N_A N_B}{1+\delta_{AB}} \frac{C_{AB} \Delta t}{S_{AB}V_{\text{cell}}} g_{AB}^{1-2\omega} \label{eq:nss}
    \end{equation}
with the characteristic constant $C_{\text{AB}}$ for the species pair $A-B$ and the VHS exponent $\omega$. Thus, a particle is restricted to collide only once per time step. Another advantage of the NSS approach is the straightforward definition of a third collision partner in a recombination reaction. However, the computational time step in the cell needs to be chosen small enough to ensure that the collision probability always remains below unity. \\
Unlike the species-specific weighting schemes developed for the CP version of the DSMC method, the collision pair selection procedure and the number of collision pairs are largely independent of the relative species weights. However, the collision probability and the update schemes of the velocity and the internal energies need to be adjusted and further steps need to be introduced for chemically reactive systems. These extensions developed in the course of this work are detailed in the following section.

\subsection{Particle Weighting} \label{ssec:vMPF} % A,B descriptions
In the following, collisions between simulation particles of species $A$ and $B$ are considered for the case $w_A < w_B$, such that a simulation particle of species $B$ represents a larger number of physical particles than a particle of species $A$. As a result, only a fraction of the real particles represented by particle $B$ can interact with the particles represented by simulation particle $A$. For the particle $A$, collision partners are available for all represented real gas particles, such that the entire simulation particle participates in the collision process and is fully influenced by it. In contrast, only a fraction of the real particles represented by particle $B$ undergoes a collision during an interaction process. In the modified collision probability, the smallest of the particle weights $\min(w_{A}, w_{B})$ now enters additionally, in order to correctly represent the fraction of real particles influenced by the collision process
 \begin{equation}
     P_{\text{c}} = \frac{w_{A} w_{B}}{\min(w_{A}, w_{B})}\frac{N_A N_B}{1+\delta_{AB}} \frac{C_{AB}\Delta t}{S_{AB}V_{\text{cell}}} g_{AB}^{1-2\omega}. \label{eq:vmpf}
 \end{equation}
The above equation reduces to Equation \ref{eq:nss} for equal species weights $w_A=w_B$. After a successful interaction, the post-collision velocity update scheme is based on the conservative weighting approach of Galitzine and Boyd \cite{boyd1996conservative,galitzine2015adaptive} for the CP model. As all real particles represented by the simulation particle of lower weight, the velocity of particle $A$ is always updated.
The velocity for the particle of species $A$, which carries the lower weight, is always updated, as the particle is fully influenced by the process. In contrast to this, the velocity of particle $B$ is updated only with a probability $P$ corresponding to the ratio of the particle weights:
\begin{equation}
    v' = \begin{cases}
        v'\hspace{0.2cm} \text{with} \hspace{0.2cm} P = \frac{\min(w_A,w_B)}{w_B}  \\
        v \hspace{0.25cm} \text{otherwise.}
    \end{cases}
\end{equation}
As a consequence, multiple successful collisions are required on average before the velocity of the higher-weight particle changes. Thus, the physically correct fraction of particles participating in a collision is reproduced on average, while additionally conserving momentum and energy. During the final sampling step of a computational time step, all particle weights are taken into account in the determination of the macroscopic averaged quantities.
% Equation
A sampling in the steady state is required to guarantee the physical accuracy of the simulation results as energy and momentum conservation can only be ensured on average in the trace weighting scheme.

\subsubsection{Trace Species Reactions}
Whenever internal energy exchange or chemical reactions are to be considered between differently weighted particles, further modifications to the standard DSMC routines are required. The modified reaction scheme is illustrated for a generic reaction of the form:
\begin{equation}
	\:A\:+\:B \rightarrow \:C \label{eq:ex_rct}
\end{equation}
In this example, all reactant and product particles have a different weighting factor. Species $A$ has the lowest weighting factor among the reactants, while the product species $C$ has the lowest overall weight. The complete reaction procedure for the species-specific weighting scheme is summarized below. A more detailed description of the chemical reaction and relaxation routines in PICLas without a trace species scheme can be found in \cite{pfeiffer2016direct,nizenkov2017modeling}. \\

1) Initially, the particle weights and properties of all reactant and product species are determined and the collision probability between particles $A$ and $B$ is evaluated according to Equation \ref{eq:vmpf}. The relative species weights of the product particles are not considered in the probability equation. If the collision probability is larger than a random number $R \in [0;1]$, the  collision event is accepted and the following steps are performed. Otherwise, no collision is taking place and the reactants remain unchanged. \\

2) As in elastic collisions, only a fraction of the colliding particles would react or relax in a physical collision process. To reproduce this behavior, an update decision is made for all reactant and product particles. If the particle weight of a given species $X$ is larger than the threshold $\min(w_A,w_B)$, the particle is updated only with the probability $P = \frac{\min(w_A,w_B)}{w_X}$. For product species with weighting factors smaller than $\min(w_A,w_B)$, additional simulation particles need to be created in a later step to preserve the correct resulting number density and ensure mass in the stochastical mean. The number of additional particles is determined according to
\begin{equation}
    N_{\text{add,X}} = \left  \lfloor \frac{\min(w_A,w_B)}{w_X} + R \right \rfloor.
\end{equation}
Thus, for each species, the average reaction rate in terms of the simulation particle number can be defined as
\begin{equation}
    \frac{\text{d}N_X}{\text{d}t} = \frac{\min(w_A,w_B)}{w_X} \frac{\text{d}N_{\min(w_A,w_B)}}{\text{d}t}.
\end{equation}

3) For simulations involving ions and electrons, using the ambipolar diffusion approximation \cite{bird1989computation}, an additional charge conservation step is required to ensure quasineutrality throughout the simulation. In this case, a uniform weighting factor is employed for all charged species and identical update decision are made.  \\

4) All product particles are subsequently created with their corresponding weighting factors. This includes products for which the update decision was negative, as they act as temporary pseudo-particles in the energy and velocity update scheme and are removed again in the final step. \\

5) The total collision energy available for redistribution is determined from the relative translational energy, by the reduced mass $\mu$ and the relative particle velocity $c_{\text{rel}}$. In addition, the internal energy of the collision partners and the reaction enthalpy $\Delta H_{\text{r}}$ are taken into consideration
\begin{align}
    E_{\text{coll}} = & \frac{1}{2}\mu c_{\text{rel}} + E_{\text{rot,A}} + E_{\text{rot,B}} + E_{\text{vib,A}} + E_{\text{vib,B}} \nonumber\\
    & + E_{\text{el,A}} + E_{\text{el,B}} - \Delta H_{\text{r}}
\end{align}
Here, the respective particle weights of the colliding species are not considered. All subsequent energy redistribution and velocity calculations are performed as if all particles possessed unit weight $w_X=1$. As the particles with a higher weight are updated only with a given probability, momentum and energy conservation are ensured on average over many collision events.\\

6 ) The available collision energy is redistributed according to the Larsen-Borgnakke method \cite{borgnakke1975statistical} among the respective internal degrees of freedom of all product particles, including the pseudo-particles. The relaxation probabilities themselves are not modified by the unequal particle weights. The remaining translational collision energy after redistribution is given by
\begin{equation}
    E_{\text{coll}}^* = E_{\text{coll}} - (E_{\text{rot,X}} + E_{\text{vib,X}} + E_{\text{el,X}}).
\end{equation}
If additional particles are to be created for trace products, the relaxation procedure is performed independently for each particle, such that independent samples are created instead of simple clones with equal energy values. The average of the determined internal energy values then enters into the above equation.\\

7) With the remainder of the collision energy, the new post-collision velocities of all product particles are determined, including the pseudo-particles. In cases with additional product samples, multiple independent velocity samples are determined. As a consequence, the number of simulation particles for trace species may increase dynamically throughout the simulation. \\

8) All reactant particles marked for update are removed from the simulation. In addition, all pseudo-product particles generated solely for the energy redistribution are deleted as well.\\

If only an internal energy relaxation is taking place, without chemical reaction processes, the same procedure is applied without the particle creation and deletion steps. In this case, the new energy and velocity values are only assigned to particles which are marked for update, while all remaining particles remain unchanged. Previous simulations have shown that the overall flow properties are less sensitive to the conservation of mass, momentum and energy in the species-specific weighting scheme, whenever particles are continuously removed and introduced in the flow.

\section{Verification}

\subsection{Reservoir Simulations}

\begin{figure}[t]
	\centering
   \includegraphics{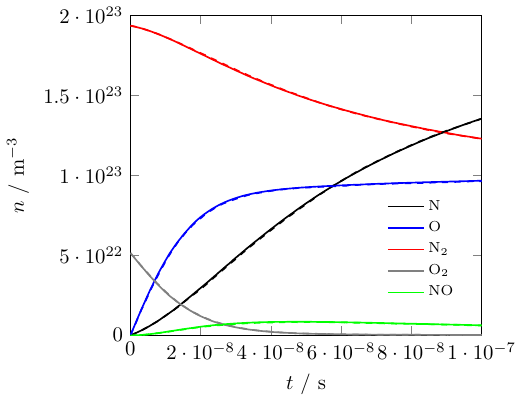}
	\caption{Temporal evolution of the simulated number density for a reactive five-species air mixture. The reference simulation with uniform weights is indicated with dashed lines, while the simulation with species-specific weights is shown with solid lines.\label{fig:res_dens}}
\end{figure}

\begin{figure}[t]
	\centering
   \includegraphics{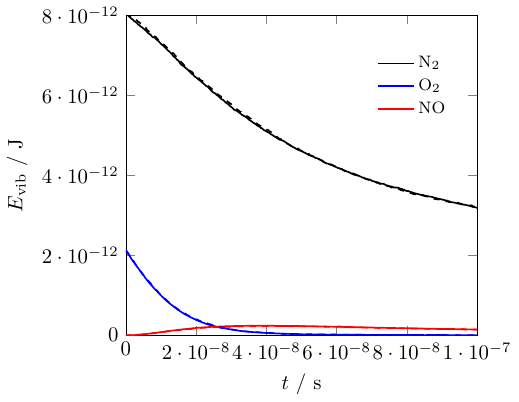}
	\caption{Temporal evolution of the simulated vibrational energy for a reactive five-species air mixture. The reference simulation with uniform weights is given with dashed lines, while the simulation with species-specific weights is shown with solid lines.\label{fig:res_en}}
\end{figure}

To verify the developed species-specific weighting scheme, a closed-system reservoir simulation of reactive air coupled to an isothermal heat bath is investigated. In contrast to open systems, non-energy conservation has a more pronounced effect in these cases, as new particles are not introduced in the simulation. Chosen for the simulation setup is a single computational cell with a volume $V= \num{1E-12}\:\mathrm{m^{3}}$. The simulation is initialized at a temperature $T=$ 30000 K, with molecular nitrogen at a density $n_{\mathrm{N_2}}= \num{1.9355E+23}\:\mathrm{m^{-3}}$ and molecular oxygen at a density $n_{\mathrm{O_2}}=\num{5.145E+022}\: \mathrm{m^{-3}}$. The gas mixture may undergo nineteen dissociation and exchange reactions, with Arrhenius reaction rates according to \cite{boyd2017nonequilibrium}. As a reference case, a uniform particle weight of $w=20$ is chosen for all occurring species. In a second simulation, the initially abundant species, molecular nitrogen and oxygen, are assigned a higher weight of $w_A=100$, while the trace species N, O and NO, formed solely through chemical reactions are assigned a lower particle weight $w_B=10$.
For both weighting cases, the simulations were run up to an end time of $t_{\mathrm{end}} =$ \num{1E-7}~s, with a computational time step of $\Delta t =$ \num{1E-10}~s. In Figure \ref{fig:res_dens}, the temporal evolution of the species number densities are compared.
The temporal evolution of the species concentrations obtained with the species-specific weighting scheme is in very good agreement with the reference simulation. In particular, the increasing concentration of the product species is reproduced with high accuracy. Thus, mass conservation and the correct reproduction of the reaction rate can be ensured with the trace species model in DSMC. To investigate energy conservation, the vibrational energies with and without the use of the species-specific particle weights are compared for the diatomic species in Figure \ref{fig:res_en}.
For both simulation cases, the vibrational energies are in close agreement with each other, independent of the relative particle weights. Thus, the reservoir simulations allowed to showcase that the species-specific weighting scheme accurately reproduces chemical reaction rates under conservation of energy on average.

\subsection{Cylinder Atmospheric Entry}
\begin{figure}[t]
    \centering
    \includegraphics{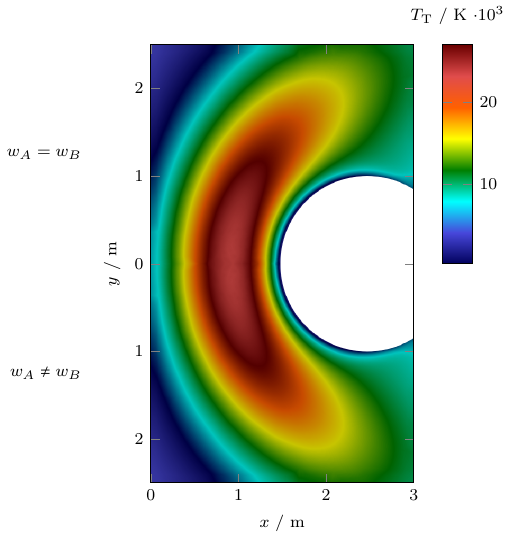}
	\caption{Total temperature flow field for the atmospheric entry of a cylinder at an altitude of 108 km, with and without the use of species-specific weighting factors. \label{fig:cyl_flow}}
\end{figure}

\begin{figure}[t]
	\centering
    \includegraphics{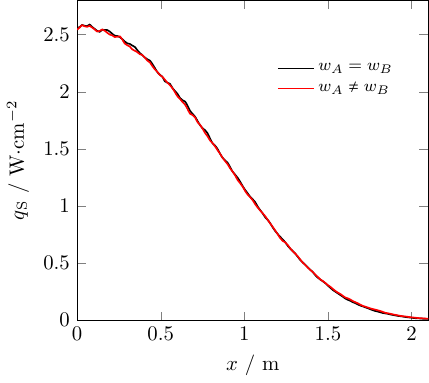}
	\caption{Comparison of the heat flux on a cylinder surface during atmospheric entry at an altitude of 108 km, with and without the use of species-specific weighting factors.\label{fig:trace_heat}}
\end{figure}

\begin{figure}
	\centering
    \includegraphics{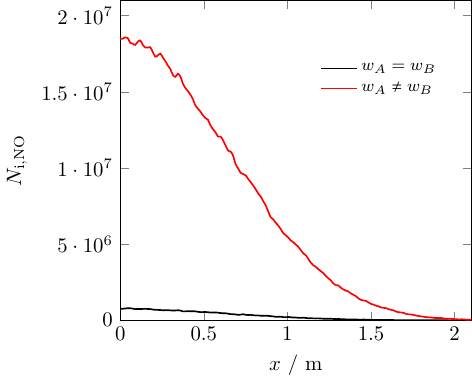}
	\caption{Comparison of the impact simulation particle number on a cylinder surface during atmospheric entry at an altitude of 108 km, with and without the use of species-specific weighting factors.\label{fig:trace_impact}}
\end{figure}

As a more sophisticated verification case, the hypersonic flow around a 2D cylinder with a radius $r=1$~m is simulated. A five-species air mixture, with nineteen possible dissociation and exchange reactions \cite{boyd2017nonequilibrium} is again considered. The free-stream conditions are chosen corresponding to an atmospheric entry trajectory at an altitude of 108 km and are summarized in Table \ref{table:trace_cylinder}.
\begin{table*}\renewcommand{\arraystretch}{1.4}
    \centering
    \caption{Free-stream conditions for the flow around a cylinder
        \cite{molchanova2018surface}.\label{table:trace_cylinder}}
    \begin{tabular}{ccccccc}
        \hline
        $Kn$ & $v_{\infty}$ / $\unit{m\cdot s^{-1}}$ & $T_{\infty}$ / K & $\rho$ / $\unit{kg\cdot m^{-3}}$ & $\chi_{\mathrm{N_2}}$ & $\chi_{\mathrm{O_2}}$ & $\chi_{\mathrm{O}}$ \\\hline
         0.65 & \num{7814.1} & 209.5 & \num{2.25E-8} & 0.771 & 0.123 & 0.106 \\\hline
    \end{tabular}
\end{table*}
In order to assess the performance of the species-specific weighting scheme, two simulations are performed for the cylinder entry. First, a uniform weighting factor $w=$ \num{1E13} is employed for all species. In the second case, a ten times lower weighting factor $w_B=$ \num{1E12} is chosen for atomic nitrogen and nitrogen monoxide, which were not present in the free-stream and are generated solely through gas-phase chemical reactions. The weighting factor for the remainder of the species is kept at $w_A=$ \num{1E13}. For the simulation with the species-specific weighting factors a time step of $\Delta t$ = \num{5E-8}~s is required in order to sufficiently resolve the collision frequency. With the use of uniform species weights, a higher time step of up to $\Delta t$ = \num{1E-7}~s can be chosen, while still resolving all relevant quality criteria, as the collision probability differs. To evaluate the accuracy of the novel weighting scheme, the temperature flow field for both simulation cases is compared in Figure \ref{fig:cyl_flow}.
The translational temperature for the reference case and for the simulation with the species-specific weighting scheme are in excellent agreement with each other.
As a further evaluation of the influence on higher-order moments of the distribution function, the heat flux along the cylinder surface for both simulation cases is compared in Figure \ref{fig:trace_heat}.
The simulation with the species-specific weighting scheme yielded a heat flux in excellent agreement with the reference case. Thus, the use of different weighting factors for the species does not result in a loss of accuracy, even in complex cases involving gas-phase chemistry. In Figure \ref{fig:trace_impact}, the overall number of impacting nitrogen monoxide simulation particles is compared.
With the use of different weighting factors, the simulation particle number representing the trace component NO can be significantly increased. However, the use of the modified collision probability, according to Equation \ref{eq:vmpf}, requires a reduced time step in order to ensure reliable results. Nevertheless, the reduced time step increased the computational cost of the simulation only by a factor of approximately two, while the sampling rate of the trace species could be tenfold improved. Achieving comparable sampling statistics with uniform particle weights would require an approximately ten times longer simulation. In this way, the overall efficiency of the simulation can be improved.
To summarize, the verification case showed a significant improvement in the statistical sampling of trace species, especially with regard to the surface properties.

\section{Results}
\begin{figure}[t]
	\centering
	\includegraphics[width=0.85\linewidth]{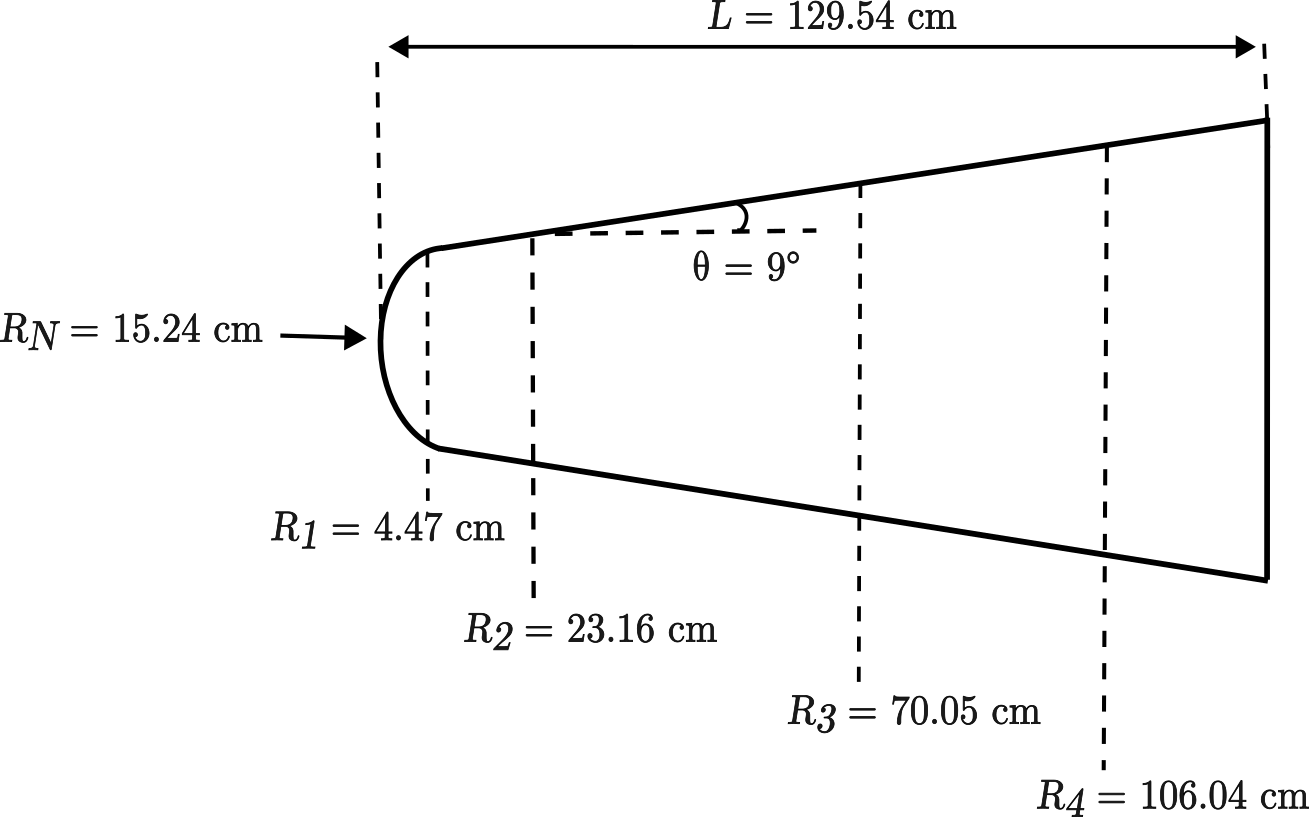}
	\caption{Schematic representation of the simulated RAM-C II capsule, including the location of the four measurement probes \cite{grantham1970flight}. \label{fig:RAM}}
\end{figure}

\begin{figure}[t]
    \centering
    \includegraphics{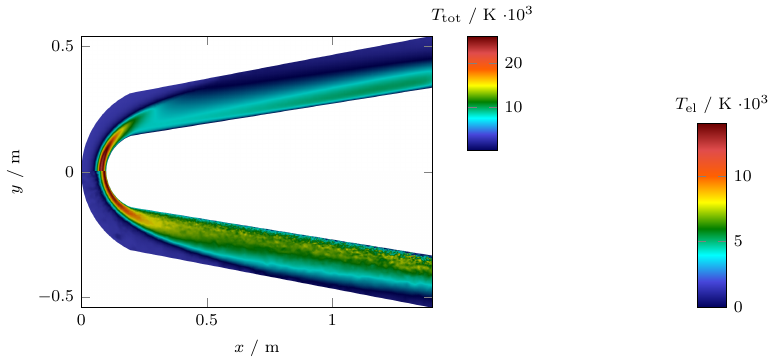}
	\caption{Total translational and electronic temperature flow field for the atmospheric entry of the RAM-C II capsule, at an altitude of 81 km. \label{fig:ram_flow}}
\end{figure}

\begin{figure}[t]
	\centering
    \includegraphics{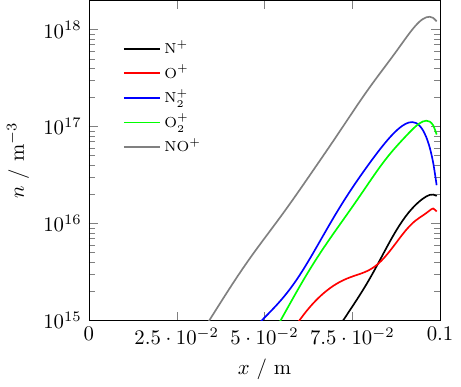}
	\caption{Cation number density evaluation for the atmospheric entry of the RAM-C II capsule at an altitude of 81 km.  \label{fig:ram_cat}}
\end{figure}

\begin{figure}[t]
	\centering
    \includegraphics{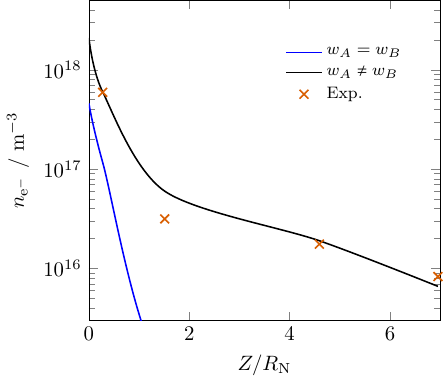}
	\caption{Comparison of the simulated electron density, with and without the use of species-
specific weighting factors, to experimental values \cite{grantham1970flight,cross1972electrostatic} for the atmospheric entry of the RAM-C II capsule at an altitude of 81 km.  \label{fig:ram_dens}}
\end{figure}

The RAM-C (Radio Attenuation Measurement for study of Communications blackout) flight experiments, conducted in the 1960s, aimed to measure the ionization rate in the compressed shock layer in the front of a spacecraft \cite{grantham1970flight,cross1972electrostatic}. Simulations of this experiment have been previously used to verify several new weighting schemes, but with the CP and not the NSS collision model \cite{fang2020dsmc,boyd2007modeling,shevyrin2020calculation}. Here, measurements of the electron density at an altitude of 81~km are investigated with the use of the novel species-specific weighting scheme in PICLas. The chosen test case represents the more rarefied part of the experimental dataset, with a Knudsen number of 0.03, based on the capsule's nose radius.
The RAM-C capsule configuration, as detailed in Figure \ref{fig:RAM}, features four probes along the surface, at which points the electron density can be obtained.
An axisymmetric 2D simulation setup is employed in order to reduce the overall computational demand. Considered is an eleven species air mixture consisting of the major neutral air species (O, N, O$_2$, N$_2$, NO) and their ionization products, together with electrons. The reaction model according to Park \cite{park1989nonequilibrium} includes both neutral reaction processes and a variety of ionization processes in the gas phase. For the capsule surface, fully diffuse reflection and full thermal accommodation is assumed, at a constant wall temperature of $T_{\text{S}}$~=~1500~K. Charged species are neutralized upon surface impact. The free-stream conditions corresponding to an entry trajectory at 81 km \cite{grantham1970flight,cross1972electrostatic}, are given in Table \ref{tab:ram_c}.
\begin{table*}\renewcommand{\arraystretch}{1.4}
	\centering
	\caption{Free-stream conditions for the atmospheric entry of the RAM-C II capsule at an altitude of 81 km \cite{grantham1970flight,cross1972electrostatic}. \label{tab:ram_c}}
	\begin{tabular}{ccccc}
		\hline
		  $Kn$ & $T_{\infty}$ / K & $v_{\infty}$ / $\unit{m\cdot s^{-1}}$ & $\rho_{\mathrm{N_2}}$ / $\unit{m^{-3}}$ & $\rho_{\mathrm{O_2}}$ / $\unit{m^{-3}}$  \\ \hline
		  0.03& 196.7 & 7800 & \num{2.587E20} &  \num{0.686E20} \\ \hline
	\end{tabular}
\end{table*}
For the charged species, the ambipolar diffusion approximation is employed. In this approximation, the electrons maintain their individual velocities while having their location fixed on the cations and move in accordance with them. Thus, quasi-neutrality is ensured throughout the flow.
To fully resolve the collision frequency, a time step of $\Delta t$ = \num{1E-10}~s is chosen in the simulation.
Two different weighting schemes are evaluated for the simulation setup. In the first instance, a uniform particle weight of $w_a = w_B =$ \num{1E11} is chosen for both the neutral and the charged species in a reference simulation. In addition, a simulation using the novel trace species weighting scheme is performed. For charged species in this case, the electrons and all cation species, the weighting factor is set to $w_A$ = \num{1E9}, while a higher weighting factor of $w_B$ = \num{1E11} is chosen for all neutral species. With this lower weighting factor, sufficient statistical sampling of all trace processes can be ensured. A uniform weighting factor is chosen for all charged species in the trace species model to ensure charge conservation in all instances.
In Figure \ref{fig:ram_flow}, the simulated total and electron translational temperature profiles are evaluated for the simulation using species-specific weighting factors.
The total translational temperature reaches values up to 26000~K in the densest part of the shock. Downstream of the shock front, the temperature is lower but still high close to the capsule surface, with values reaching up to 12000~K. Ionization processes are favored in regions of higher temperature. Thus, the ionization degree closely aligns with the temperature distribution. The temperature of the created electrons is in non-equilibrium with the remainder of the flow, with values up to 14000~K in the shock front. Again, the electron temperature decreases along the surface to a value of 6000~K. \\
In Figure \ref{fig:ram_cat}, relative concentrations of the cation species are evaluated for the simulation using the trace species model.
The formation of the nitrogen monoxide cation is favored, as the molecule most effectively stabilizes the positive charge. Atomic cations are disfavored under these conditions, as reflected by the higher ionization energy.\\
Figure~\ref{fig:ram_dens} compares the simulated electron density along the capsule surface to the simulation using uniform particle weights and to experimental flight measurement data.
The electron density is highest in the shock layer, in the region of the highest overall temperature, and it decreases along the surface line.  With the use of uniform particle weights in the DSMC simulation, the experimental electron densities cannot be reproduced, as the charged trace species are severely underrepresented in this case. For sufficient resolution of the electrons, all species weighting factors would need to be increased by at least a factor of 100. As this would accordingly increase the computational demand, a simulation with that many particles would be cumbersome. By the use of the trace species model on the other hand, the simulated electron density is in good agreement with the measured flight data for all probe locations. Even though a lower weighting factors is employed for the electrons in the simulation using the species-specific weighting scheme, the total simulation particle number decreases significantly along the surface line. Thus, the statistical noise apparent in the results increased as well. Therefore, the ionization degree in the regions at the back of the capsule might not be fully represented and may deviate from the experimental dataset. Additionally, the limitations of the ambipolar diffusion approximation may contribute to the observed differences, especially at the probes 2 and 4. Overall, the results demonstrate that the DSMC module of PICLas, with the ambipolar diffusion approximation and trace species weighting, can accurately capture ionization processes during atmospheric entry in the given test case. However, to further improve the accuracy in the electron description, a fully coupled PIC-DSMC simulation should be performed.

\section{Conclusion}
A species-specific weighting scheme has been implemented into the open-source particle code PICLas. The weighting scheme specially considered reactive processes under conservation of mass, momentum and energy. The implementation was verified using reservoir simulations of a reactive five-species air mixture.
Furthermore, the surface heat flux and sampling statistics for the flow around a 2D cylinder were compared to a reference simulation with uniform species weights. With the novel weighting scheme, the sampling for the trace species could be significantly improved without loss of accuracy. In addition, the ionization degree along an atmospheric entry trajectory was investigated and compared with RAM-C II flight measurement data, where the electron density showed good agreement with the experimental data.\\
For future applications, the efficiency of DSMC simulations can be further improved by coupling the species-specific weighting scheme with additional adaptive methods. With the extension of the approach to include spatially varying particle weights, the resolution of large density gradients in the simulation domain can be significantly improved. In addition, the use of a variable time step could reduce the impact of the reduced time step required by the modified collision probability, which would further improve the overall computational performance.

\section*{Acknowledgments}
	This work is supported by the Deutsche Forschungsgemeinschaft project number 516238647 - SFB1667/1 (ATLAS - Advancing Technologies for Low-Altitude Satellites). The authors gratefully acknowledge the computing time provided on the high-performance computer HoreKa by the National High-Performance Computing Center at KIT (NHR@KIT). This center is jointly supported by the Federal Ministry of Research, Technology and Space and the Ministry of Science, Research and the Arts of Baden-Württemberg, as part of the National High-Performance Computing (NHR) joint funding program. HoreKa is partly funded by the German Research Foundation (DFG).

% ----------------------------------------------------------------------------------------------- %
% Data Availability Statement
% ----------------------------------------------------------------------------------------------- %
\section*{Data Availability Statement}
The data that support the findings of this study are available from the corresponding author upon reasonable request.

% ----------------------------------------------------------------------------------------------- %
% Bibliography
% ----------------------------------------------------------------------------------------------- %
\bibliographystyle{elsarticle-num}
\bibliography{vMPF.bib}

@book{bird1994molecular,
		title={Molecular gas dynamics and the direct simulation of gas flows},
		author={Bird, Graeme A},
		year={1994},
		publisher={Oxford university press}
}

@article{baganoff1990collision,
  title={A collision-selection rule for a particle simulation method suited to vector computers},
  author={Baganoff, D and McDonald, JD},
  journal={Physics of Fluids A: Fluid Dynamics},
  volume={2},
  number={7},
  pages={1248--1259},
  year={1990},
  publisher={American Institute of Physics}
}

@book{mcdonald1990computationally,
  title={A computationally efficient particle simulation method suited to vector computer architectures},
  author={McDonald, Jeffrey Douglas},
  year={1990},
  publisher={Stanford University}
}

@article{fasoulas2019combining,
  title={Combining particle-in-cell and direct simulation Monte Carlo for the simulation of reactive plasma flows},
  author={Fasoulas, Stefanos and Munz, C-D and Pfeiffer, Marcel and Beyer, Julian and Binder, Tilman and Copplestone, Stephen and Mirza, Asim and Nizenkov, Paul and Ortwein, Philip and Reschke, Wladimir},
  journal={Physics of Fluids},
  volume={31},
  number={7},
  year={2019},
  publisher={AIP Publishing}
}

@article{galitzine2015adaptive,
  title={An adaptive procedure for the numerical parameters of a particle simulation},
  author={Galitzine, Cyril and Boyd, Iain D},
  journal={Journal of Computational Physics},
  volume={281},
  pages={449--472},
  year={2015},
  publisher={Elsevier}
}

@article{pfeiffer2013grid,
  title={A grid-independent particle pairing strategy for DSMC},
  author={Pfeiffer, Marcel and Mirza, Asim and Fasoulas, Stefanos},
  journal={Journal of Computational Physics},
  volume={246},
  pages={28--36},
  year={2013},
  publisher={Elsevier}
}

@article{boyd1996conservative,
  title={Conservative species weighting scheme for the direct simulation Monte Carlo method},
  author={Boyd, Iain D},
  journal={Journal of Thermophysics and Heat Transfer},
  volume={10},
  number={4},
  pages={579--585},
  year={1996}
}

@book{grantham1970flight,
  title={Flight results of a 25000-foot-per-second reentry experiment using microwave reflectometers to measure plasma electron density and standoff distance},
  author={Grantham, William L},
  volume={6062},
  year={1970},
  publisher={National Aeronautics and Space Administration}
}

@book{cross1972electrostatic,
  title={Electrostatic-probe Measurements of Plasma Parameters for Two Reentry Flight Experiments at 25 000 Feet Per Second},
  author={Cross, Aubrey E},
  year={1972},
  publisher={NASA}
}

@inproceedings{bird1989computation,
  title={Computation of electron density in high altitude re-entry flows},
  author={Bird, G},
  booktitle={20th Fluid Dynamics, Plasma Dynamics and Lasers Conference},
  pages={1882},
  year={1989}
}

@article{takahashi2016analysis,
  title={Analysis of radio frequency blackout for a blunt-body capsule in atmospheric reentry missions},
  author={Takahashi, Yusuke and Nakasato, Reo and Oshima, Nobuyuki},
  journal={Aerospace},
  volume={3},
  number={1},
  pages={2},
  year={2016},
  publisher={MDPI}
}

@article{park1989nonequilibrium,
  title={Nonequilibrium hypersonic aerothermodynamics},
  author={Park, Chul},
  journal={Nonequilibrium hypersonic aerothermodynamics by Park},
  pages={29860},
  year={1989}
}

@article{borgnakke1975statistical,
  title={Statistical collision model for Monte Carlo simulation of polyatomic gas mixture},
  author={Borgnakke, Claus and Larsen, Poul S},
  journal={Journal of computational Physics},
  volume={18},
  number={4},
  pages={405--420},
  year={1975},
  publisher={Elsevier}
}

@book{boyd2017nonequilibrium,
  title={Nonequilibrium gas dynamics and molecular simulation},
  author={Boyd, Iain D and Schwartzentruber, Thomas E},
  volume={42},
  year={2017},
  publisher={Cambridge University Press}
}

@article{molchanova2018surface,
title={Surface recombination in the direct simulation Monte Carlo method},
  author={Molchanova, Alexandra N and Kashkovsky, Alexander V and Bondar, Yevgeniy A},
  journal={Phys. Fluids},
  volume={30},
  number={10},
  pages={107105},
  year={2018},
  publisher={AIP Publishing LLC}
}

@article{pfeiffer2016direct,
  title={Direct simulation Monte Carlo modeling of relaxation processes in polyatomic gases},
  author={Pfeiffer, M and Nizenkov, P and Mirza, A and Fasoulas, S},
  journal={Physics of Fluids},
  volume={28},
  number={2},
  year={2016},
  publisher={AIP Publishing}
}

@article{nizenkov2017modeling,
  title={Modeling of chemical reactions between polyatomic molecules for atmospheric entry simulations with direct simulation Monte Carlo},
  author={Nizenkov, P and Pfeiffer, M and Mirza, A and Fasoulas, S},
  journal={Physics of Fluids},
  volume={29},
  number={7},
  year={2017},
  publisher={AIP Publishing}
}

@article{fang2020dsmc,
  title={DSMC modeling of rarefied ionization reactions and applications to hypervelocity spacecraft reentry flows},
  author={Fang, Ming and Li, Zhi-Hui and Li, Zhong-Hua and Liang, Jie and Zhang, Yong-Hao},
  journal={Advances in Aerodynamics},
  volume={2},
  number={1},
  pages={7},
  year={2020},
  publisher={Springer}
}

@article{boyd2007modeling,
  title={Modeling of associative ionization reactions in hypersonic rarefied flows},
  author={Boyd, Iain D},
  journal={Physics of Fluids},
  volume={19},
  number={9},
  year={2007},
  publisher={AIP Publishing}
}

@article{shevyrin2020calculation,
  title={On the calculation of the electron temperature flowfield in the DSMC studies of ionized re-entry flows},
  author={Shevyrin, Alexander and Bondar, Yevgeniy},
  journal={Advances in Aerodynamics},
  volume={2},
  number={1},
  pages={6},
  year={2020},
  publisher={Springer}
}

\end{document}